\documentclass[%
 aip,
 apl,
 amsmath,amssymb,
 reprint,%
]{revtex4-2}
\usepackage{float}
\usepackage{textcomp}
\usepackage{blindtext}
\usepackage{chemformula}
\usepackage{graphicx}
\usepackage{dcolumn}
\usepackage{bm}

\usepackage{changes}

\usepackage{xcolor}
\usepackage{xr-hyper}
\usepackage{hyperref}
\usepackage[T1]{fontenc}
\usepackage{mathptmx}

\begin{document}

\preprint{AIP/APL}

\title[Study of charge density waves in suspended 2H-\ch{TaS2} and 2H-\ch{TaSe2} by nanomechanical resonance]{Study of charge density waves in suspended 2H-\ch{TaS2} and 2H-\ch{TaSe2} by nanomechanical resonance}

\author{Martin Lee}
\email[]{M.Lee-2@tudelft.nl}
\thanks{These two authors contributed equally.}
\affiliation{Kavli Institute of Nanoscience, Delft University of Technology, Lorentzweg 1, 2628 CJ Delft, The Netherlands.}
\author{Makars \v{S}i\v{s}kins}
\thanks{These two authors contributed equally.}
\affiliation{Kavli Institute of Nanoscience, Delft University of Technology, Lorentzweg 1, 2628 CJ Delft, The Netherlands.}
\author{Samuel Ma\~{n}as-Valero}
\affiliation{Instituto de Ciencia Molecular (ICMol) Universitat de Val\`{e}ncia, c/ Catedr\'{a}tico Jos\'{e} Beltr\'{a}n 2, 46980 Paterna, Spain.}
\author{Eugenio Coronado}
\affiliation{Instituto de Ciencia Molecular (ICMol) Universitat de Val\`{e}ncia, c/ Catedr\'{a}tico Jos\'{e} Beltr\'{a}n 2, 46980 Paterna, Spain.}
\author{Peter G. Steeneken}
\affiliation{Kavli Institute of Nanoscience, Delft University of Technology, Lorentzweg 1, 2628 CJ Delft, The Netherlands.}
\affiliation{Department of Precision and Microsystems Engineering, Delft University of Technology, Mekelweg 2, 2628 CD Delft, The Netherlands.}
\author{Herre S. J. van der Zant}
\affiliation{Kavli Institute of Nanoscience, Delft University of Technology, Lorentzweg 1, 2628 CJ Delft, The Netherlands.}

\date{\today}
\begin{abstract}

The charge density wave (CDW) state in van der Waals systems shows interesting scaling phenomena as the number of layers can significantly affect the CDW transition temperature, $T$\textsubscript{CDW}. However, it is often difficult to use conventional methods to study the phase transition in these systems due to their small size and sensitivity to degradation. Degradation is an important parameter which has been shown to greatly influence the superconductivity in layered systems. Since the CDW state competes with the onset of superconductivity, it is expected that $T$\textsubscript{CDW} will also be affected by the degradation. Here, we probe the CDW phase transition by the mechanical resonances of suspended 2H-\ch{TaS2} and 2H-\ch{TaSe2} membranes and study the effect of disorder on the CDW state. Pristine flakes show the transition near the reported values of 75 K and 122 K respectively. We then study the effect of degradation on 2H-\ch{TaS2} which displays an enhancement of $T$\textsubscript{CDW} up to 129 K after degradation in ambient air. Finally, we study a sample with local degradation and observe that multiple phase transitions occur at 87 K, 103 K and 118 K with a hysteresis in temperature in the same membrane. The observed spatial variations in the Raman spectra suggest that variations in crystal structure cause domains with different transition temperatures which could result in the hysteresis. This work shows the potential of using nanomechanical resonance to characterize the CDW in suspended 2D materials and demonstrate that degradation can have a large effect on transition temperatures.

\end{abstract}

\maketitle%

The charge density wave (CDW) state in van der Waals (vdW) materials has recently become a resurgent area of research. Archetypal systems such as 2H-\ch{NbSe2}, 2H-\ch{TaS2} and 2H-\ch{TaSe2} have been under study since the 1970s \cite{wilson1975charge, tsang1976raman, sugai1985lattice, sugai1981studies, harper1975heat, craven1977specific, delaplace1976on,rice1975new, campagnoli1979plasmon, steignmeier1976softening, Tsang1977raman, mcmillan1976theory, moncton1977neutron, wilson1978questions}. However, recent works on surprising and unexpected layer dependence and degradation effects on superconductivity (SC) and CDW in these systems have revived interest in studying their phase transitions. For example, the superconducting transition temperature, $T$\textsubscript{SC} of 2H-\ch{NbSe2} is suppressed from 7.2 K in the bulk to 3 K in the monolayer limit while the CDW transition temperature, $T$\textsubscript{CDW} is increased\cite{xi2015strongly} from 33 K to 145 K. More surprisingly, 2H-\ch{TaS2} has a $T$\textsubscript{SC} of 0.6 K in the bulk which increases to 3 K in a monolayer \cite{sergio2018tuning,navarro2016enhanced} and a $T$\textsubscript{CDW} of 75 K which also increases to 140 K \cite{zhang2020charge}. Similar scaling is seen for $T$\textsubscript{SC} in 2H-\ch{TaSe2}\cite{wu2018dimensional}. Furthermore, degradation of the crystal in air has shown to enhance the superconductivity in 2H-\ch{TaS2} \cite{bekaert2020enhanced} which is in stark contrast to other air-sensitive vdW superconductors which lose their superconductivity upon degradation \cite{yang2018stability,novoselov2005two,sandilands2014origin,yu2019high}. It is an ongoing challenge to clarify these contradicting layer dependencies and degradation effects in order to shine light on the competition between CDW and SC in these materials.

The CDW transition, like other first and second order phase transitions, can be described by Landau's theory of phase transitions \cite{landau1968statistical} where the emergence of charge order gives rise to a sudden change in the specific heat. Using the specific heat anomaly to probe the phase transition is already established in several systems \cite{craven1977specific, nyeanchi1994specific, takano2004magnetic,testardi1975elastic, loram2004absence, grabovsky2013calorimetric}. However, the traditional methods of probing the specific heat are nearly impossible to apply on ultrathin exfoliated 2D material flakes. Recent works on using the nanomechanical resonance to extract various phase transitions including structural, magnetic and electronic phase transitions \cite{vsivskins2020magnetic, davidovikj2020ultrathin, jiang2020exchange, sengupta2010electromechanical} have shown to be an interesting alternative.

In this work, we study the CDW transitions of suspended 2H-\ch{TaS2} and 2H-\ch{TaSe2} flakes by tracking the temperature dependence of their nanomechanical resonance frequency. The resonance frequency of suspended pristine 2H-\ch{TaS2} and 2H-\ch{TaSe2} flakes show an anomaly at the phase transition temperatures of 75 K and 122 K respectively. We then employ this technique as a probe to study the effect of degradation on the $T$\textsubscript{CDW} of 2H-\ch{TaS2}. Flakes of 2H-\ch{TaS2} show greatly enhanced $T$\textsubscript{CDW} after being exposed to ambient conditions for prolonged durations. Furthermore, we induce local disorder in a region of a suspended part of the membrane which causes varying degrees of disorder across the flake as observed in Raman spectroscopy. In this sample, multiple transitions appear with a hysteretic switching behavior pointing towards the existence of domains with varying $T$\textsubscript{CDW}.

\begin{figure}[ht]
\includegraphics[width=\columnwidth]{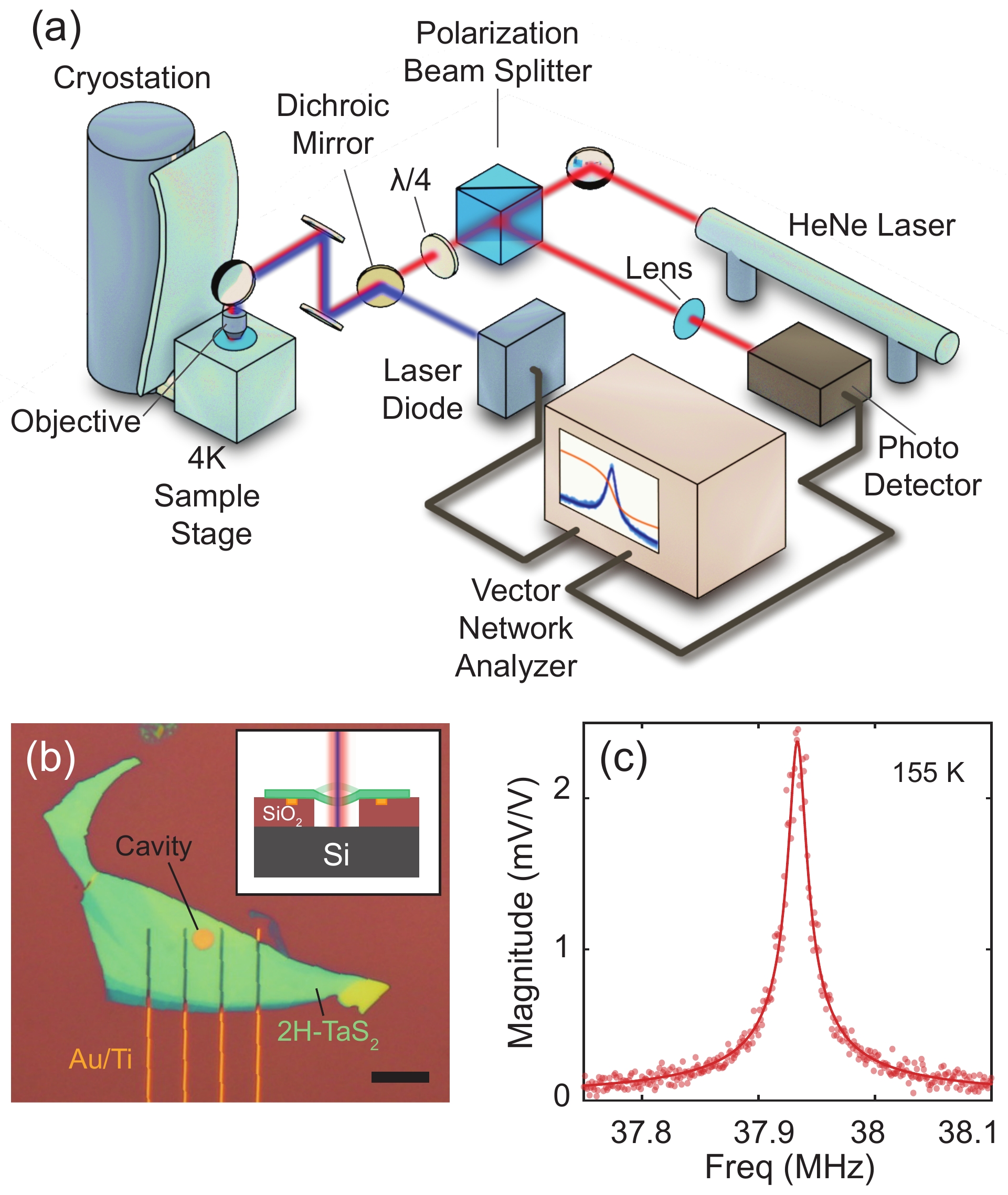}
\caption{Interferometry setup, device geometry and basic characterization of a 2H-\ch{TaS2} membrane. \textbf{(a)} Illustration of the laser interferometry setup. Blue diode laser is used to optothermally actuate the membrane while the He-Ne red laser is used to read out its motion. \textbf{(b)} Optical image of device 1 composed of a 2H-\ch{TaS2} flake of thickness $t$ = 31.2$\pm$0.6 nm transferred onto pre-defined electrodes surrounding a cavity. Scale bar: 10 \textmu m. Inset: Illustration of the cross-section of a device. \textbf{(c)} Example frequency response of device 1 at 155 K.}
\label{fig:setup}
\end{figure}

The interferometry setup and the sample are described in Fig. \ref{fig:setup}. Figure \ref{fig:setup}(a) shows an illustration of the interferometry setup. The intensity of the blue diode laser ($\lambda$\textsubscript{Blue} = 405 nm) is modulated by the vector network analyzer (VNA) which optothermally excites the membrane into motion. Simultaneously, a continuous He-Ne laser ($\lambda$\textsubscript{Red} = 632 nm) is used to read out the movement of the membrane. The interference signal is collected by the photodetector which is read out by the VNA. The sample is situated in a 4 K dry cryostat at high vacuum with a heater beneath the sample to control the temperature.

Devices are fabricated by deterministically stamping \cite{castellanos2014deterministic} 2H-\ch{TaS2} and 2H-\ch{TaSe2} flakes on top of electrodes metallized by evaporation and circular cavities etched into \ch{SiO2/Si} by reactive ion etching. The suspended membrane is in a drum geometry with a rigid Si back mirror. High-quality 2H-\ch{TaS2} and 2H-\ch{TaSe2} flakes are exfoliated from synthetically grown bulk crystals \cite{navarro2016enhanced,freitas2016strong}. Detailed description of the setup and the fabrication processes can be found in S.I.

An optical image of device 1 is shown in Fig. \ref{fig:setup}(b), the cross-sectional illustration is in the inset and its typical frequency response at 155 K near the fundamental resonance frequency in Fig. \ref{fig:setup}(c). The data is collected at every temperature once stabilized to within 10 mK from the set-point and fitted to a simple harmonic oscillator model. The fundamental resonance frequency $f_0(T)$, extracted from such sweeps is plotted in frequency vs. temperature plots in subsequent figures.

In this section, we introduce the lambda type anomaly in the specific heat due to the normal - CDW phase transition, as described by Landau-Lifshitz \cite{landau1968statistical,saint2019survey,vsivskins2020magnetic}. The Landau free energy can be written for CDW transitions as:
\begin{equation}
F=F_0+a(T-T_\textrm{CDW})Q^2+BQ^4,
\label{eq:landau}
\end{equation}
where $F_0$ is the temperature dependent free energy of the normal state, $Q$ is the order parameter and $a$ and $B$ are phenomenological positive constants. Minimizing the above equation with respect to $Q$ (i.e., $\partial F / \partial Q$ = 0), gives the CDW order parameter:
\begin{equation}
Q=\sqrt{\frac{-a(T-T_\textrm{CDW})}{2B}},
\label{eq:orderparameter}
\end{equation}
and a minimum free energy $F_\textrm{min}=F_0-\frac{a^2}{4B}(T-T_\textrm{CDW})^2$. Using the relation for the specific heat at constant pressure, $c_\textrm{p}(T)=-T\left[\frac{\partial^2 F}{\partial T^2}\right]_\textrm{P}$ and by substituting the expression for $F_\textrm{min}$ into Eq. \ref{eq:landau}, the magnitude of the jump in the specific heat at the phase transition can be derived as $\Delta c_\textrm{p}= \frac{a^2T_\textrm{CDW}}{2B}$. To find the relationship between the membrane resonance frequency and $c_\textrm{p}(T)$ we note that the fundamental resonance frequency of a circular membrane under thermal strain can be described by:

\begin{equation}
f_0(T)=\frac{2.4048}{\pi d}\sqrt{\frac{E}{\rho} \frac{\epsilon(T)}{(1-\nu)}},
\label{eq:freq}
\end{equation}
where $d$ is the membrane diameter, $E$ the Young's modulus, $\rho$ the density, $\epsilon(T)$ the temperature dependent biaxial strain, and $\nu$ Poisson's ratio. 

The thermal strain accumulated in the membrane is a result of the difference in the linear thermal expansion coefficient of the membrane $\alpha_\textrm{L}$ and that of the substrate $\alpha_\textrm{Si}$. It can be expressed as: $\frac{d\epsilon(T)}{dT}\simeq -(\alpha_\textrm{L}(T)-\alpha_\textrm{Si}(T))$ assuming that the thermal expansion coefficient of \ch{SiO2} is negligible in comparison to Si \cite{white1977thermal,lyon1977linear}. Using the thermodynamic relation between the thermal expansion coefficient and the specific heat, $\alpha_\textrm{L}(T)=\gamma c_\textrm{v}(T)/(3KV_\textrm{M})$, and the above-mentioned thermal strain relation, we arrive at an expression:

\begin{equation}
c_\textrm{v}(T)=3\left(\alpha_\textrm{si}-\frac{1}{\mu^2}\frac{d[f^2_0(T)]}{dT}\right)\frac{KV_\textrm{M}}{\gamma},
\label{eq:cv}
\end{equation}
where $c_\textrm{v}(T)$ is the specific heat of the membrane at constant volume, $K=\frac{E}{3(1-2\nu)}$ the bulk modulus, $\gamma$ the Gr{\"u}neisen parameter, $V_\textrm{M}=M/\rho$ the molar volume of the membrane and $\mu=\frac{2.4048}{\pi d}\sqrt{\frac{E}{\rho (1-\nu)}}$ a constant. We note that the Young's modulus $E$ is also slightly temperature dependent and exhibits an anomaly at the phase transition. However, this change is on the order of a percent throughout the temperature range of our experiment \cite{barmatz1975elasticity}. We therefore approximate it as a constant and assume the thermal strain to be dominant in determining the frequency changes \cite{vsivskins2020magnetic}. Since $c_\textrm{v}\simeq c_\textrm{p}$ in solids, Eq. \ref{eq:cv} directly relates the mechanical resonance of the membrane to the specific heat derived using the Landau free energy (Eq. \ref{eq:landau}). Through this relation we can extract the specific heat from the temperature derivative of $f_0^2$ and thus the $T$\textsubscript{CDW} from determining the discontinuity in the specific heat. Detailed discussion on this relation can be found in Ref.\cite{vsivskins2020magnetic} and the supplementary materials S.II.

\begin{figure}
\includegraphics[width=\columnwidth]{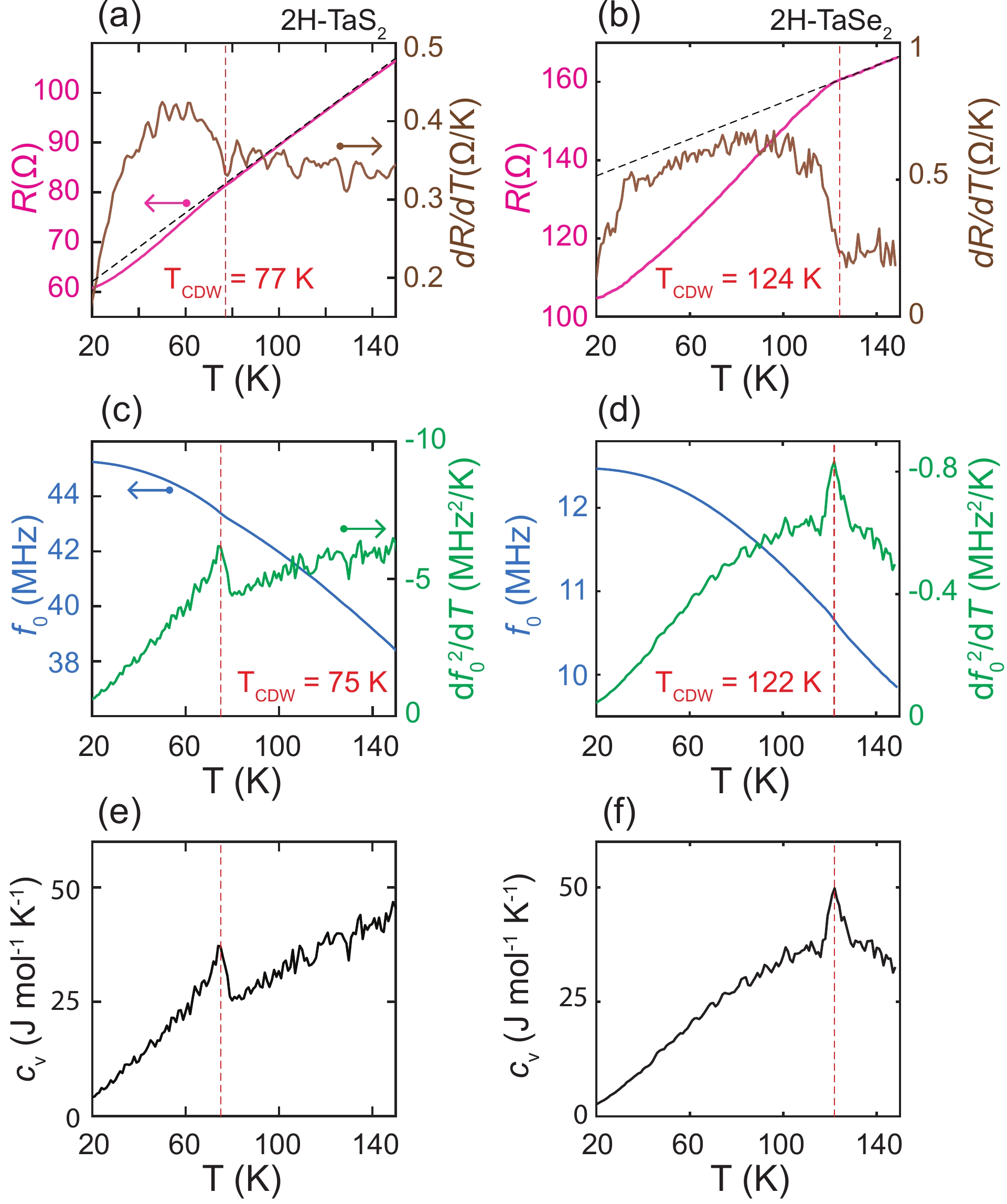}
\caption{Electrical and mechanical characterization of a pristine 2H-\ch{TaS2} (left column, device 1) and 2H-\ch{TaSe2} (right column, device 2) membrane. Dashed red lines indicate the CDW transition temperature determined from $\frac{dR}{dT}$ and the peak of $\frac{d[f^2_0(T)]}{dT}$. \textbf{(a-b)} Four-probe resistance as a function of temperature (left y-axis, pink) and its derivative (right y-axis, brown). Dashed straight black lines are plotted as visual aid. \textbf{(c-d)} Resonance frequency $f_0$ (left y-axis, blue) and $\frac{d[f^2_0(T)]}{dT}$ (right y-axis, green).  \textbf{(e-f)} Specific heat extracted from (c-d) using Eq. \ref{eq:cv}. The data from device 1 is also used in Ref. \cite{vsivskins2020magnetic}, \v{S}i\v{s}kins et al, Nature communications, Vol. 11, Article 2698, 2020; licensed under a Creative Commons Attribution (CC BY) license.}
\label{fig:standard}
\end{figure}

\begin{figure}
\includegraphics[width=\columnwidth]{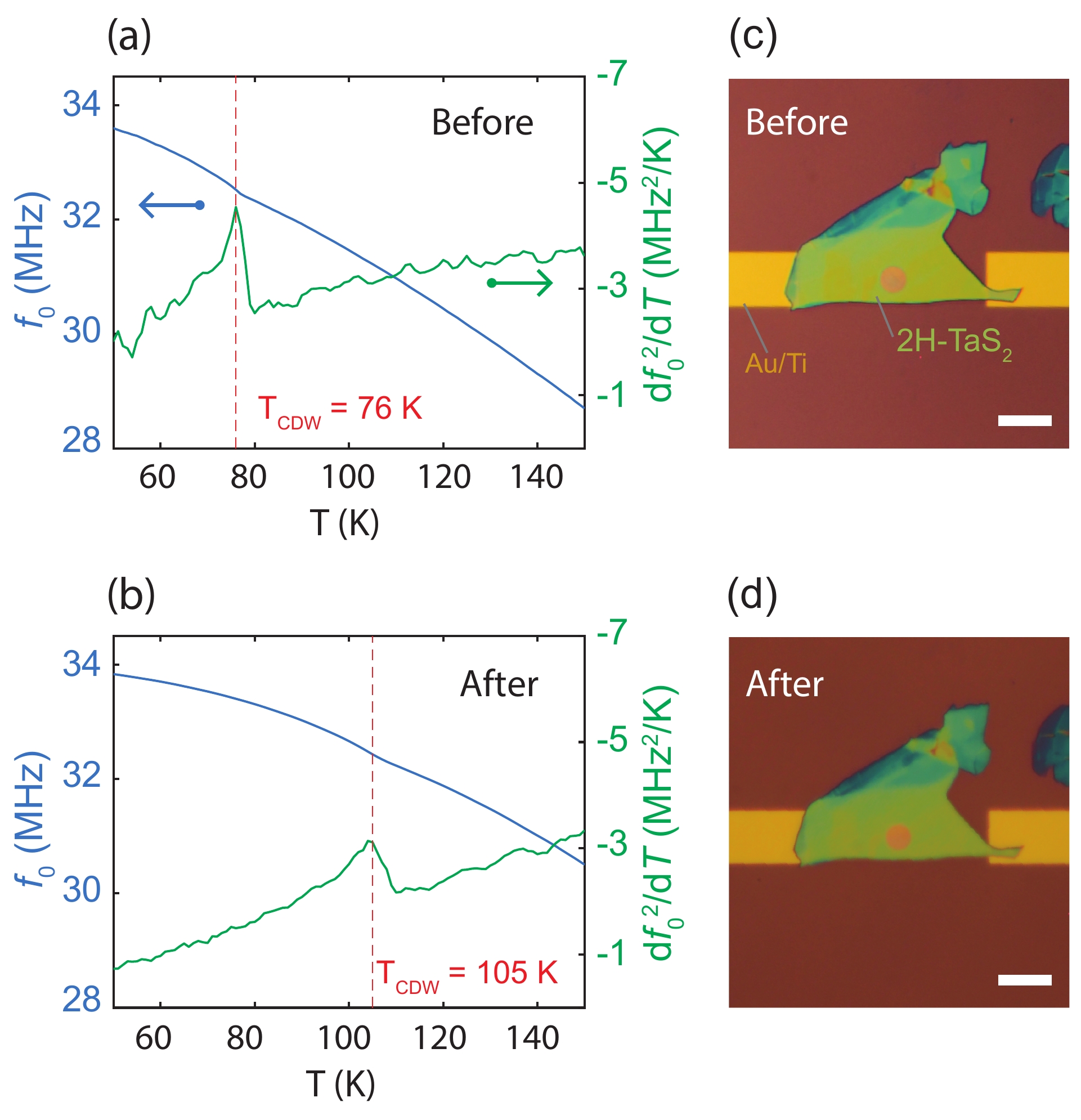}
\caption{Enhancement of the $T$\textsubscript{CDW} from the pristine state to the degraded state in device 3 (2H-\ch{TaS2}). $f_0$ (blue, left y-axis) and $\frac{d[f^2_0(T)]}{dT}$ (green, right y-axis) measured \textbf{(a)} immediately following fabrication and \textbf{(b)} after exposure to ambient conditions for several hours. Optical image of the device \textbf{(c)} immediately after fabrication and \textbf{(d)} after measurements of (a-b). Scale bar: 10 \textmu m. Dashed red lines correspond to the $T$\textsubscript{CDW}.}
\label{fig:enhancement}
\end{figure}

\begin{figure*}[ht!]
\includegraphics[width=0.99\textwidth]{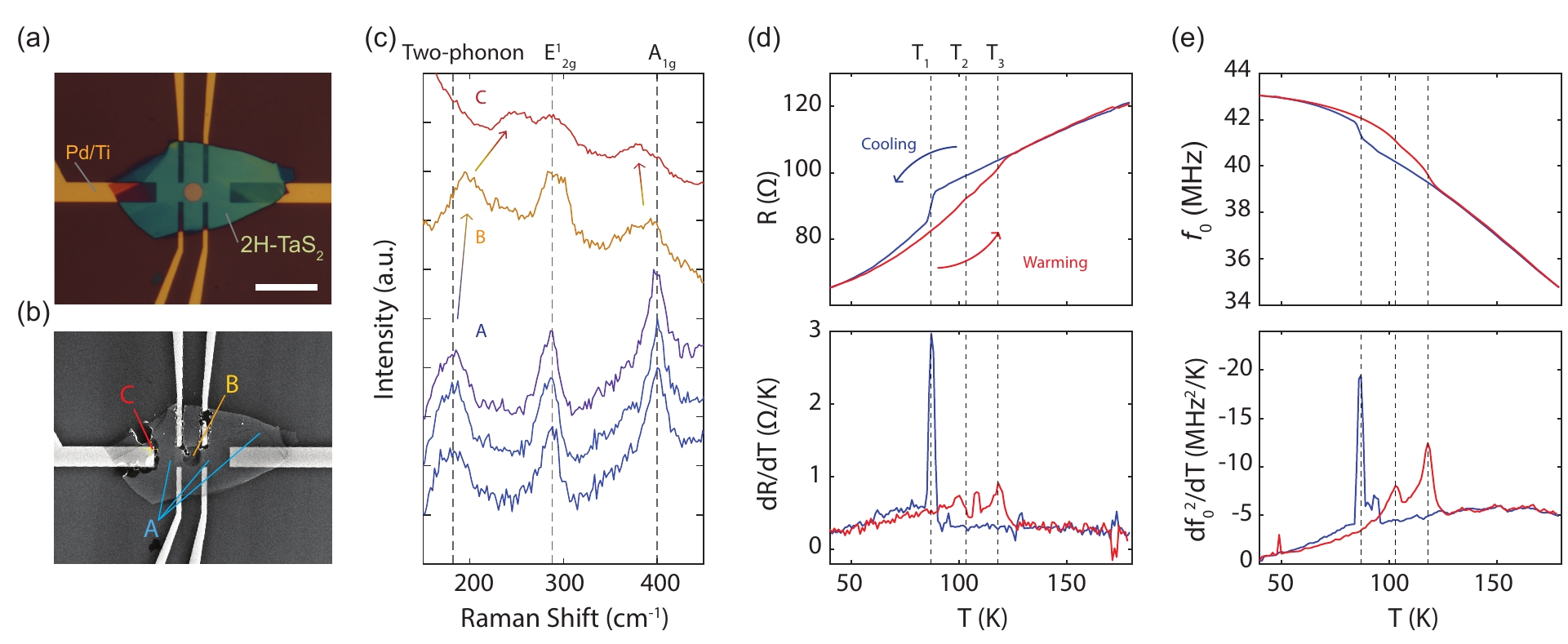}
\caption{Raman, electrical and mechanical characterizations of device 4 (2H-\ch{TaS2}) showing competing transitions of CDW with enhanced $T$\textsubscript{CDW}. \textbf{(a)} Optical image of device 4 immediately after the fabrication. Scale bar: 10 \textmu m. \textbf{(b)} SEM image of device 4 after measurement. Labels A, B and C indicate the positions where Raman spectroscopy data were acquired. \textbf{(c)} Raman spectroscopy data at A, B and C. Dashed lines indicate the position of two-phonon mode, E$^1_{2g}$ and A$_{1g}$. \textbf{(d)} Two-point resistance as a function of temperature (top) and its derivative (bottom). \textbf{(e)} Resonance frequency $f_0$, as a function of temperature (top) and $\frac{d[f^2_0(T)]}{dT}$ (bottom). In both (d) and (e), the blue lines indicate measurements taken while cooling down and red lines indicate measurements taken while warming up as indicated by the arrows in (d). Dashed lines indicate the positions of the transition temperatures $T_{1}$ = 87 K, $T_{2}$ = 103 K, $T_{3}$ = 118 K.}
\label{fig:hysteresis}
\end{figure*}

Temperature dependent mechanical and electrical responses of pristine flakes of 2H-\ch{TaS2} (left column, device 1, thickness, $t$ = 31.2 $\pm$ 0.6 nm, $d$ = 4 \textmu m) and 2H-\ch{TaSe2} (right column, device 2, $t$ = 23.3 $\pm$ 0.5 nm , $d$ = 10 \textmu m, device not shown) are shown in Fig. \ref{fig:standard}. Figure \ref{fig:standard}(a-b) show the four-probe resistance (left y-axis, pink) and its derivative (right y-axis, brown). Dashed black lines are plotted as a visual guide to highlight the deviation of the resistance data from the linear drop. The kink below which the resistance drop deviates from the dashed black line is the CDW transition temperature universally seen in other CDW systems \cite{lin2020patterns}. This can also be seen in the $\frac{dR}{dT}$ as the temperature at which the slope changes. The CDW transition temperatures for 2H-\ch{TaS2} and 2H-\ch{TaSe2} by analysing the $\frac{dR}{dT}$ are 77 K and 122 K respectively and are in good agreement with the values in literature \cite{sugai1985lattice}.

On the same membranes, the resonance frequency $f_0$ extracted by fitting a simple harmonic oscillator function to the resonance peak such as in Fig. \ref{fig:setup}(c) is plotted against temperature in Fig. \ref{fig:standard}(c-d) in blue (left y-axis). There is a monotonic increase in the resonance frequency as the sample temperature is lowered, arising from the difference in the thermal expansion coefficient between the membrane and the substrate thus increasing the tension of the resonator. The temperature derivative of $f_0^2$ is plotted in green in Fig. \ref{fig:standard}(c-d) (right y-axis). Since $c_\textrm{v}\propto\frac{d[f_0^2(T)]}{dT}$ from Eq. \ref{eq:cv}, the phase transition temperature $T$\textsubscript{CDW} can be determined as the temperature where the peak in $\frac{d[f^2_0(T)]}{dT}$ is observed. The $T$\textsubscript{CDW} of 2H-\ch{TaS2} and 2H-\ch{TaSe2} extracted from Fig. \ref{fig:standard}(c-d) are 75 K and 122 K respectively, and are in good agreement with the values from the transport data as well as the literature values \cite{sugai1985lattice}.Therefore, this method can be a complementary tool to the transport technique to probe the phase transition in CDW materials which show subtle changes in the slope of the resistance. In the subsequent sections, the $T$\textsubscript{CDW}'s are extracted by finding the peak position of the anomaly in $\frac{d[f^2_0(T)]}{dT}$ vs. $T$.

The $c_\textrm{v}(T)$ can be estimated from the same data by including the material parameters into Eq. \ref{eq:cv}. The reported material parameters for 2H-\ch{TaS2} are $E = E_\textrm{2D}/t=149$ GPa \cite{jiang2017parameterization} assuming an interlayer spacing $t=0.58$ nm, $\nu = 0.27$ \cite{jiang2017parameterization} and $\rho=6110$ kg/m$^{3}$. The parameters for 2H-\ch{TaSe2} are $E = 120$ GPa \cite{barmatz1975elasticity}, $\nu\sim 0.2$ \cite{kang2013band} and $\rho=8660$ kg/m$^{3}$. The Gr{\"u}neisen parameters can be estimated as $\gamma\simeq\frac{3}{2}(\frac{1+\nu}{2-3\nu})$ \cite{sanditov2011relation}. Finally, the temperature dependent thermal expansion coefficient of single crystalline Si is used as experimentally measured in Ref. \cite{middelmann2015thermal}. Using these parameters, the $\frac{d[f^2_0(T)]}{dT}$ data in Fig. \ref{fig:standard}(c-d) are converted to $c_\textrm{v}$ and plotted in Fig. \ref{fig:standard}(e-f).

For the remainder of the study, we focus on the effect of degradation on the CDW transition, specifically in 2H-\ch{TaS2}. We first use the above-mentioned technique to probe the phase transition temperature in 2H-\ch{TaS2} before and after prolonged exposure to air. Degradation is often accompanied by changes in the material properties such as doping \cite{bekaert2020enhanced}, Poisson's ratio \cite{hao2018oxidation}, Young's modulus \cite{falin2021mechanical}, dimensions \cite{li2019recent} and density\cite{lavik1968oxidation}. Therefore, in the following, we extract the transition temperatures from $\frac{d[f^2_0(T)]}{dT}$ plots and refrain from showing $c_\textrm{v}$ in order to circumvent errors in $c_\textrm{v}$ arising from using wrong material parameters in Eq. \ref{eq:cv}.

To study degradation effects on the CDW transition temperature, device 3 ($t$ = 53.3 $\pm$ 0.7 nm, $d$ = 5 \textmu m) is measured before and after exposure to ambient conditions. In the first measurement, it is cooled down immediately following fabrication. Figure \ref{fig:enhancement}(a) shows $f_0$ (blue) as well as the temperature derivative of $f_0^2$ (green). As expected, the CDW transition occurs at $T$\textsubscript{CDW} = 76 K (dashed red line) which is in good agreement with device 1 and literature values \cite{sugai1985lattice}. After the first cool down, the sample is removed from the cryostat and is exposed to air for several hours.

In the second cool down a remarkable 29 K enhancement of the $T$\textsubscript{CDW} is observed. As shown in Fig. \ref{fig:enhancement}(b), the anomaly in $\frac{d[f^2_0(T)]}{dT}$ occurs at 105 K instead of 76 K. Several additional samples of air-degraded 2H-\ch{TaS2} have been measured, one of which showed an even higher $T$\textsubscript{CDW} of 129 K (see supplementary material S.III). In contrast to the drastic change in the $T$\textsubscript{CDW}, no observable changes in the optical microscopy images before and after could be identified. Figure \ref{fig:enhancement}(c) is an image of device 3 immediately after the stamping process whereas Fig. \ref{fig:enhancement}(d) is the image taken after the second round of measurements.

We have also fabricated and measured several different samples of 2H-\ch{TaS2} drums with disorder created by laser induced oxidation \cite{cartamil2015high} and focused ion beam (FIB) induced milling \cite{roslon2020high} to intentionally degrade the suspended flakes. Neither of these samples with different forms of disorder showed the CDW transition (see supplementary material S.IV).  However, in device 4, an electrostatic discharge across two electrodes adjacent to the suspended membrane caused a severe degradation of the membrane between the two electrodes. Figure \ref{fig:hysteresis}(a) shows an image of the device immediately after fabrication showing no signs of damage. In Fig. \ref{fig:hysteresis}(b), a scanning electron microscopy (SEM) image is shown of the device after the measurements are taken. It shows that the discharge caused severe damage to the the top electrodes as well as a small part of the membrane. The areas labeled A, B and C are the locations where the Raman spectroscopy data in Fig. \ref{fig:hysteresis}(c) are taken. Raman spectroscopy is performed at room temperature in ambient conditions.

The blue Raman spectra shown in Fig. \ref{fig:hysteresis}(c) are from the areas surrounding the drum and show the spectra comparable to literature \cite{navarro2016enhanced,zhang2019effects}. The three characteristic peaks of 2H-\ch{TaS2} are plotted in dashed grey lines at 180 cm$^{-1}$, 286 cm$^{-1}$ and 400 cm$^{-1}$ corresponding to the two-phonon mode, the in-plane E$^1_\textrm{2g}$ mode and the out-of-plane A$_\textrm{1g}$ mode respectively \cite{sugai1981studies}. The yellow line in Fig. \ref{fig:hysteresis}(c) is the Raman spectrum taken directly on the drum and shows slight red shifting of the two-phonon mode. The red line is the spectrum taken from the area with the most damage observed. This spectrum shows the most severely red shifted two-phonon mode as well as slightly blue shifted A$_{1g}$ mode as indicated by the arrows.

Figure \ref{fig:hysteresis}(d) shows the two-probe resistance - measured across the two wide electrodes far left and right of the cavity - of this device measured as a function of temperature (top) and its temperature derivative (bottom). The mechanical resonance of the membrane as a function of temperature (top) as well as $\frac{d[f_0]^2}{dT}$ (bottom) are plotted in Fig. \ref{fig:hysteresis}(e). In both resistance and mechanics measurements, more than one phase transition accompanied by a hysteretic behavior in the temperature sweeps are observed. The red lines correspond to the measurements performed while warming up and the blue lines to the measurements performed while cooling down. There are three distinct peaks in the  $\frac{d[f^2_0(T)]}{dT}$ at $T_{1}$ = 87 K, $T_{2}$ = 103 K and $T_{3}$ = 118 K. Between the lowest transition temperature and the highest, both the resistance and the mechanical resonance show hysteretic behavior of split branches in the R-T and the $f_0$-T data. Even though the effective area probed via transport and nanomechanics are not identical, similar behaviors are observed in both R-T and $f_0$-T suggesting that the degraded area has a significant contribution to the resistance as well as the mechanics. This experiment has been repeated multiple times to rule out measurement artifacts but nonetheless, the hysteresis was present every time.

We believe that both the thickness and degradation are playing a role in our observation of enhanced $T$\textsubscript{CDW} in Figs. \ref{fig:enhancement}\&\ref{fig:hysteresis}. The study by Bekaert et al. \cite{bekaert2020enhanced} on the ``healing'' of the sulphur vacancies by oxygen, demonstrated that the electron-phonon coupling could be enhanced by 80\%, thus increasing the $T$\textsubscript{SC}. Also, Zhang et al. recently reported the persistence of the CDW up to 140 K in the monolayer\cite{zhang2020charge}. The increase of $T$\textsubscript{CDW} upto 129 K in our air degraded sample could be an indication of partial amorphization of the multilayer sample which reduces the effective thickness of the crystal from bulk towards an intermediate, few-effective-layers.

The two-phonon mode shown in Fig. \ref{fig:hysteresis}(c) represents a second order scattering process where an electron scatters to create a pair of phonons with opposite momenta near the CDW wave vector q\textsubscript{CDW} \cite{sugai1985lattice}. Softening of the two-phonon peaks below $T$\textsubscript{CDW} in many 2H-\ch{MX2} systems has been observed and used to characterize the CDW  \cite{sugai1981studies,hajiyev2013contrast,zhang2019effects,joshi2019short,pandey2020electron,klein1981theory}. Typically the position of the two-phonon mode shifts down with decreasing temperature and the peak disappears as it reaches the CDW state. This is a direct result of the phonon dispersion renormalization due to the Kohn anomaly forming at $T$\textsubscript{CDW}. The fact that we see differences in the two-phonon mode and the out of plane A$_\textrm{1g}$ mode in the degraded areas is indicative of local changes in the phonon branches and the chemical bond lengths caused by degradation. Controlled systematic Raman study of degradation dynamics should be conducted to correlate the changes in the chemical bonds to the phonon dispersion relation and the $T$\textsubscript{CDW}.

The enhancement of the $T$\textsubscript{CDW} from the nominal 75 K up to 118 K in Fig. \ref{fig:hysteresis} may be due to a degradation similar to the one observed in Fig. \ref{fig:enhancement} but is attributed in this case to the discharge which caused the flake and the electrodes to be damaged. The absence of the peaks at 103 K and 118 K in the downward sweep and at 87 K in the upward sweep may be an indication of competition between various domains with different transition temperatures. This picture is further supported by the difference in the Raman spectra taken at room temperature in various areas of the same flake.

In conclusion, we studied the CDW transitions in the archetypal vdW systems 2H-\ch{TaS2} and 2H-\ch{TaSe2}, by using the resonance frequency of suspended membranes. The temperature dependence of the resonance frequency can be translated into the specific heat which shows an anomaly at the phase transition temperature. We showed that degradation can irreversibly change the CDW transition temperature from the nominal value of $T$\textsubscript{CDW} = 75 K to as high as 129 K. Furthermore, we studied a suspended drum with partial local disorder which showed multiple transition temperatures as well as a hysteresis loop. In this work, we have demonstrated that nanomechanical resonance is a powerful tool to study the CDW transitions in ultrathin suspended vdW materials complementary to the temperature dependent electronic transport.

\vspace{5mm}

See supplementary material for the S.I Methods, S.II Derivation of free energy and specific heat, S.III Additional data on enhancement of CDW by exposure to air and S.IV Additional methods of inducing disorder explored.
\vspace{5mm}

M.L., M.\v{S}., P.G.S. and  H.S.J.v.d.Z. acknowledge funding from the European Union's Horizon 2020 research and innovation program under grant agreement number 881603. E.C. and S.M.-V. thank the financial support from the European Union (ERC AdG Mol-2D  788222), the Spanish MICINN (MAT2017-89993-R and Excellence Unit ``Mar\'{i}a de Maeztu'', CEX2019-000919-M), and the Generalitat Valenciana (PO FEDER Program, ref. IDIFEDER/2018/061 and PROMETEO).

\vspace{5mm}

The data that support the findings of this study are openly available in Zenodo at http://doi.org/10.5281/zenodo.4719865.

\bibliography{enhancedTcdw_corrected}
\bibliographystyle{ieeetr}

\end{document}


\renewcommand{\theequation}{S.\arabic{equation}}
\renewcommand{\thefigure}{S\arabic{figure}}


\title{Supplementary Information for : \\ Study of charge density waves in suspended 2H-\ch{TaS2} and 2H-\ch{TaSe2} by nanomechanical resonance}

\author{Martin Lee}
\thanks{These two authors contributed equally.}
\affiliation{Kavli Institute of Nanoscience, Delft University of Technology, Lorentzweg 1, 2628 CJ Delft, The Netherlands.}
\author{Makars \v{S}i\v{s}kins}
\thanks{These two authors contributed equally.}
\affiliation{Kavli Institute of Nanoscience, Delft University of Technology, Lorentzweg 1, 2628 CJ Delft, The Netherlands.}
\author{Samuel Ma\~{n}as-Valero}
\affiliation{Instituto de Ciencia Molecular (ICMol) Universitat de Val\`{e}ncia, c/ Catedr\'{a}tico Jos\'{e} Beltr\'{a}n 2, 46980 Paterna, Spain.}
\author{Eugenio Coronado}
\affiliation{Instituto de Ciencia Molecular (ICMol) Universitat de Val\`{e}ncia, c/ Catedr\'{a}tico Jos\'{e} Beltr\'{a}n 2, 46980 Paterna, Spain.}
%
\author{Peter G. Steeneken}
\affiliation{Kavli Institute of Nanoscience, Delft University of Technology, Lorentzweg 1, 2628 CJ Delft, The Netherlands.}
\affiliation{Department of Precision and Microsystems Engineering, Delft University of
Technology, Mekelweg 2, 2628 CD Delft, The Netherlands.}
\author{Herre S. J. van der Zant}
\affiliation{Kavli Institute of Nanoscience, Delft University of Technology, Lorentzweg 1, 2628 CJ Delft, The Netherlands.}

\maketitle
%
%
%
\pagebreak


\section{Methods}
\label{supp:methods}

\subsection{Prepatterend \ch{SiO2/Si}}

Dry thermal oxide of 285 nm, grown on highly doped (Si++) silicon is used as the substrate. Using standard e-beam lithography (EBL), electrodes are patterned into a PMMA 495 k - 950 k bilayer. After development, exposed \ch{SiO2} areas are briefly dry etched using \ch{CHF3} and \ch{Ar} plasma in an anisotropic reactive ion etcher (RIE) such that 100 nm of the 285 nm \ch{SiO2} is removed. Using an e-beam evaporator, 5 nm Ti and 95 nm Au are evaporated in the etched structure, embedding the electrodes into the \ch{SiO2}. The wafer is submerged in acetone for lift off and rinsed in isopropanol. In the second step, circular cavities are defined using EBL and AR-P 6200 resist. After development, exposed \ch{SiO2} areas are dry etched completely down to the Si using RIE. AR-P 6200 is stripped in PRS-3000 and the sample is plasma cleaned in an \ch{O2} barrel asher prior to stamping.

\subsection{Transfer of 2H-\ch{TaS2} and 2H-\ch{TaSe2}}

The exfoliation and transfer of multi-layer 2H-\ch{TaS2} and 2H-\ch{TaSe2} flakes is done using PDMS transfer method \cite{castellanos2014deterministic}. First, PDMS is made by mixing Sylgard 184 base with the curing agent in a 10:1 ratio by mass and desiccating to remove pockets of gas. The mixture is left to cure for at least 48 hours before use. Using magic tape, 2H-\ch{TaS2} and 2H-\ch{TaS2} are exfoliated onto the PDMS. Flakes of tens of nanometers in thickness - confirmed by the optical contrast - are identified and transferred onto the set of electrodes and cavity in \ch{SiO2}/Si.

\subsection{Laser interferometry}

Sample is mounted on a heater stage which is cooled down to 4 K using a dry cryostat with optical access to the sample space. An AC signal sent from the vector network analyzer (VNA) drives the intensity of the blue diode laser ($\lambda_\textrm{blue}$ = 405 nm). The laser is focused on the center of the membrane which is optothermally driven into motion. The motion of the membrane across the optical field of a second laser (continuous red laser of $\lambda_\textrm{red}$ = 632 nm) causes an interference with the red light reflected from the Si cavity bottom and is collected at the photodetector and read by the VNA in a homodyne fashion.

\subsection{R vs. T measurement}

Temperature dependence of the sample resistance is performed using a Keysight B2902A precision source measure unit. At every temperature, current-voltage trace is measured while current biasing and the resistance is extracted by fitting a linear slope.

\subsection{Raman spectroscopy and SEM}

Raman spectroscopy is performed at room temperature using a Renishaw InVia system with a 514 nm green excitation laser. 0.5\% of 50 mW is used for the collection of the Raman data.  Scanning electron microscopy (SEM) is performed using a FEI Helios G4 CX system at 20 kV acceleration voltage.

\newpage

\section{Derivation of the change in specific heat}

\label{supp:landau}

In this section, we derive the Landau-Lifshitz \cite{landau1968statistical,saint2019survey,vsivskins2020magnetic} expression for the  charge density wave (CDW) order parameter, minimum free energy and the change in the specific heat $\Delta c_\textsubscript{v}$ at the transition temperature $T$\textsubscript{CDW}. The Landau free energy is written as:
\begin{equation}
F=F_0+a(T-T_\textsubscript{CDW})Q^2+BQ^4,
\label{seq:landau}
\end{equation}
where $F_0$ is the temperature dependent free energy of the normal state, $Q$ is the order parameter and $a$ and $B$ are phenomenological positive constants. Minimizing the free energy (Eq. \ref{seq:landau}) with respect to $Q$ by setting the derivative equal to zero, we get:
$0=2a(T-T_\textsubscript{CDW})Q+4BQ^3$
so that 
$Q^2=\frac{-a(T-T_\textsubscript{CDW})}{2B}$,
which can be written as:
\begin{equation}
Q=\sqrt{\frac{-a(T-T\textsubscript{CDW})}{2B}}.
\label{seq:orderparameter}
\end{equation}
Using Eq. \ref{seq:landau} and Eq. \ref{seq:orderparameter}, the equilibrium free energy $F_\textsubscript{min}$ equals:
\begin{equation}
F_\textsubscript{min}=F_0-\frac{a^2(T-T_\textsubscript{CDW})^2}{4B}.
\end{equation}
We can now calculate the entropy, $S=-\partial F/\partial T$ as
\begin{align*}
S_\textsubscript{min}=-\frac{\partial F_\textsubscript{min}}{\partial T}&= \frac{a^2(T-T_\textsubscript{CDW})}{B}.
\end{align*}
The difference in entropy above and below the transition temperature is:
\begin{equation}
S_\textsubscript{min}-S_0= 
\begin{cases}
\frac{a^2(T-T_\textsubscript{CDW})}{2B}& T<T_\textsubscript{CDW},\\
0 &T>T_\textsubscript{CDW},
\end{cases}
\label{seq:entropy}
\end{equation}
where $S_0$ is the entropy of the normal state.

From Eq. \ref{seq:entropy}, the specific heat at constant pressure by $c_\textsubscript{p}=T\left[\frac{\partial S}{\partial T}\right]_\textsubscript{P}$ can be found by subtracting the specific heat of the normal state. The difference in the specific heat caused by the phase transition is thus:

\begin{equation}
\Delta c_\textsubscript{p}=c_\textsubscript{p,min}-c_\textsubscript{p0}= 
\begin{cases}
\frac{a^2}{2B}& T<T_\textsubscript{CDW},\\
0 &T>T_\textsubscript{CDW}.
\end{cases}
\label{seq:cv}
\end{equation}

Since the specific heat at constant volume $c_\textsubscript{v}$ is comparable to $c_\textsubscript{p}$ (i.e., $c_\textsubscript{v}\simeq c_\textsubscript{p}$) in incompressible solids we can relate Eq. \ref{seq:cv} to Eq. 4 of the main text. The change in $c_\textsubscript{v}$ is basically the height of the anomaly - commonly refered to as the ``lambda anomaly'' due to its shape - visible in the $c_\textsubscript{v}$ vs. temperature plots in the main text Fig. 2(e-f). Compared to the data in Ref. \cite{vsivskins2020magnetic}, an improved analysis method is used to analyze the data shown in Fig. 2 of main text which may cause slight differences in the magnitude of the $c_\textsubscript{v}$ in these plots with respect to the original plots. The transition temperature values remain unaffected.

\newpage
\section{Additional sets of measurements on degradation}
\label{supp:additional}

In this section, we show additional measurements performed on multiple flakes of air degraded 2H-\ch{TaS2} stamped on two separate substrates (A and B). 

\subsection{Two flakes on substrate A}

Two devices (A1 and A2) of 2H-\ch{TaS2} on substrate A are prepared separately by cleaving freshly from the bulk crystal. The two flakes are stamped on the same substrate at relatively the same time and therefore have been exposed to the ambient conditions for nearly the same amount of time; the time exposed to the ambient is a few hours.

\begin{figure}[h!]
\includegraphics[width=0.5\columnwidth]{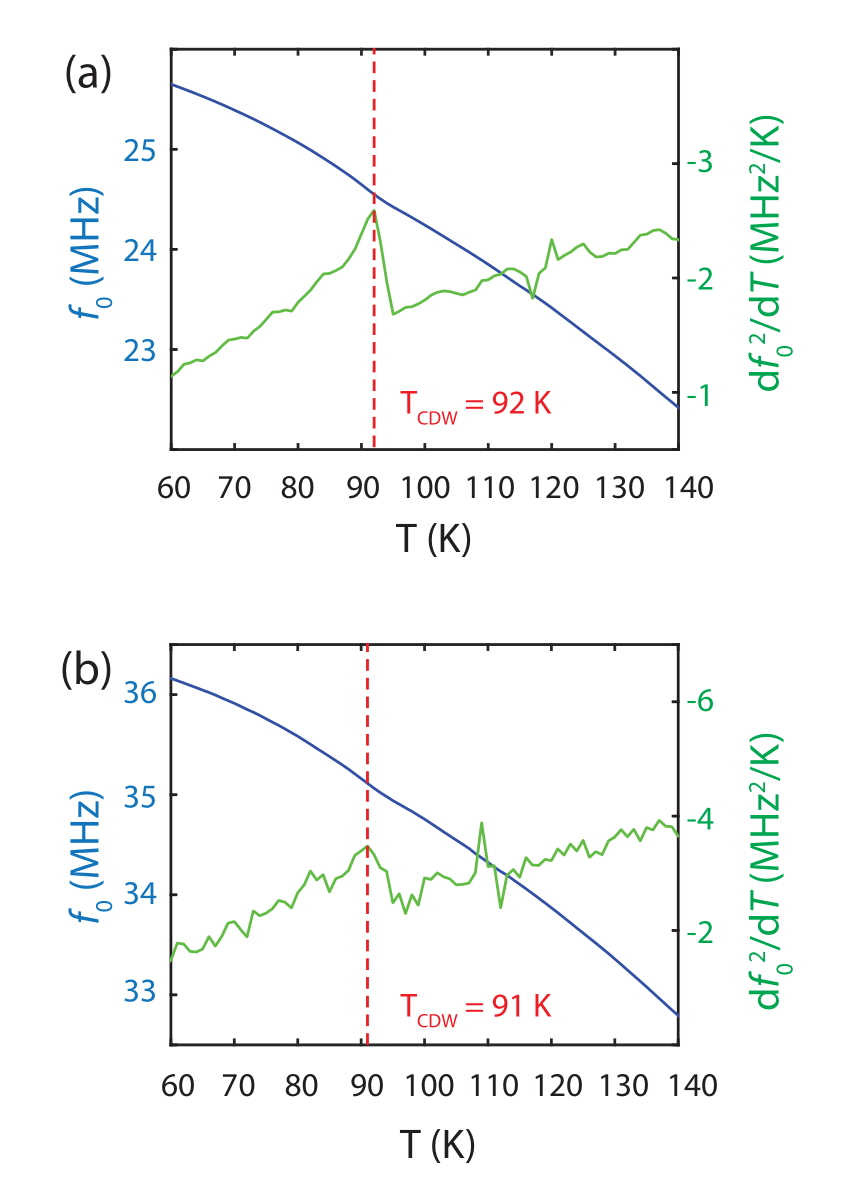}
\caption{Measurements of two flakes on substrate A. The resonance frequencies $f_0$ are plotted in blue (left y-axis) and its temperature derivative $\frac{d[f^2_0]}{dT}$ in green (right y-axis) \textbf{(a)} Device A1, showing $T$\textsubscript{CDW} = 92 K. \textbf{(b)}. Device A2 on the same substrate as device A1, showing $T$\textsubscript{CDW} = 91 K. }
\label{fig:S1}
\end{figure}

Figure \ref{fig:S1}(a-b) show the resonance frequencies $f_0$ in blue (left y-axis) and $\frac{d[f^2_0]}{dT}$ in green (right y-axis) of samples A1 and A2. Both flakes on substrate A show a similar $T$\textsubscript{CDW} of 92 K and 91 K as determined from the peaks of $\frac{d[f^2_0]}{dT}$. This demonstrates that degradation globally affected both samples on the substrate.

\subsection{Two flakes on substrate B}

\begin{figure}[h!]
\includegraphics[width=0.5\columnwidth]{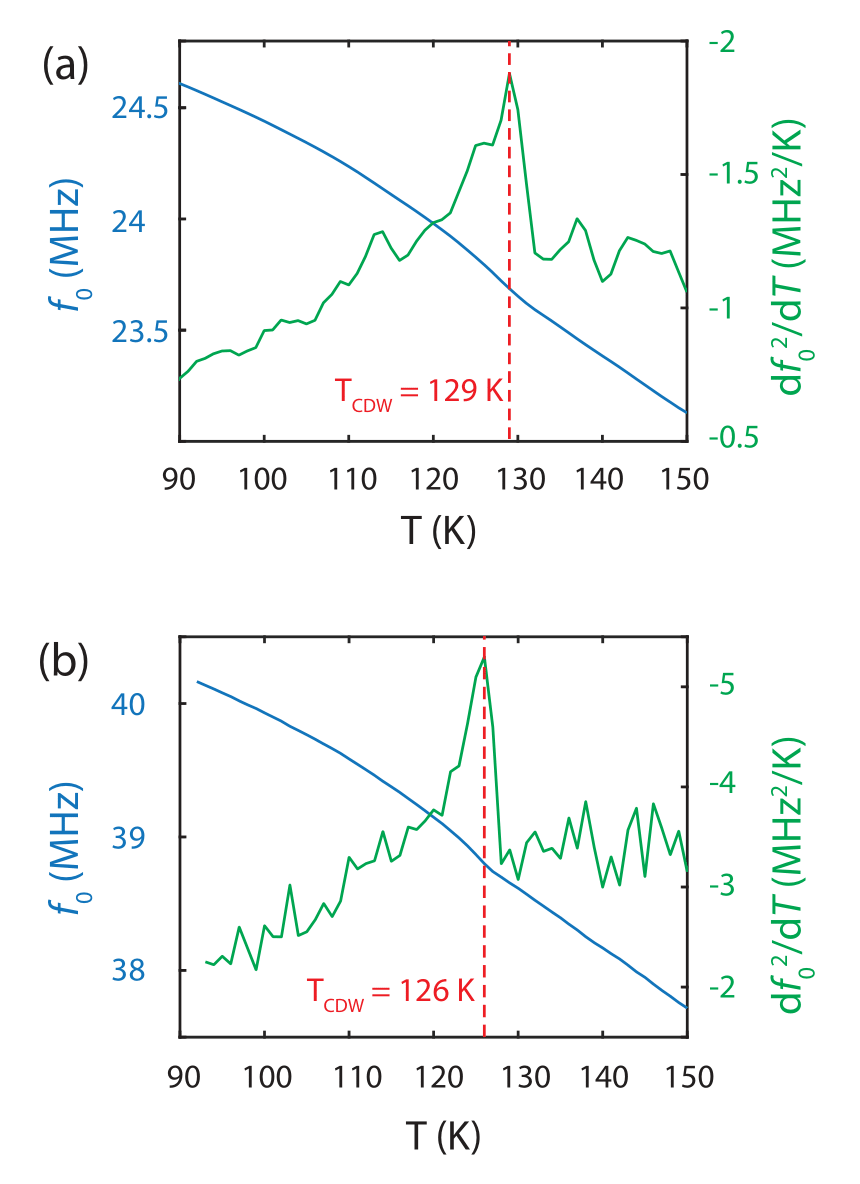}
\caption{Measurements of two flakes on substrate B. The resonance frequencies $f_0$ are plotted in blue (left y-axis) and its temperature derivative $\frac{d[f^2_0]}{dT}$ in green (right y-axis) \textbf{(a)} Device B1, showing the $T$\textsubscript{CDW} = 129 K. \textbf{(b)}. Device B2, showing the $T$\textsubscript{CDW} = 126 K. }
\label{fig:S2}
\end{figure}

A similar procedure was applied to two devices (B1 and B2) on substrate B. As shown in Fig. \ref{fig:S2}, CDW transitions occur at much higher temperatures of 126 K and 129 K.

The observation that the respective flakes on A and B show the same trend in the increased $T$\textsubscript{CDW}, indicates that the enhancement in CDW transition temperature is caused by an external factor such as air and/or humidity, rather than fluctuation in the flake quality. As far as we could observe, there were no differences in the sample preparation between samples on A and B. We suspect that the temperature and humidity variation of the day may have played a role.

\section{Other methods of inducing disorder}
\label{supp:otherdisorder}

In this section, we explore two alternative methods of inducing disorder: laser induced oxidation performed similarly to Ref. \cite{cartamil2015high} on a flake stamped on substrate C and focused ion beam (FIB) induced milling performed similarly to Ref. \cite{mykkanen2020enhancement,roslon2020high,celebi2014ultimate} on a flake stamped on substrate D.

\subsection{Laser induced oxidation on a flake on substrate C}

\begin{figure}[!h]
\includegraphics[width=0.8\columnwidth]{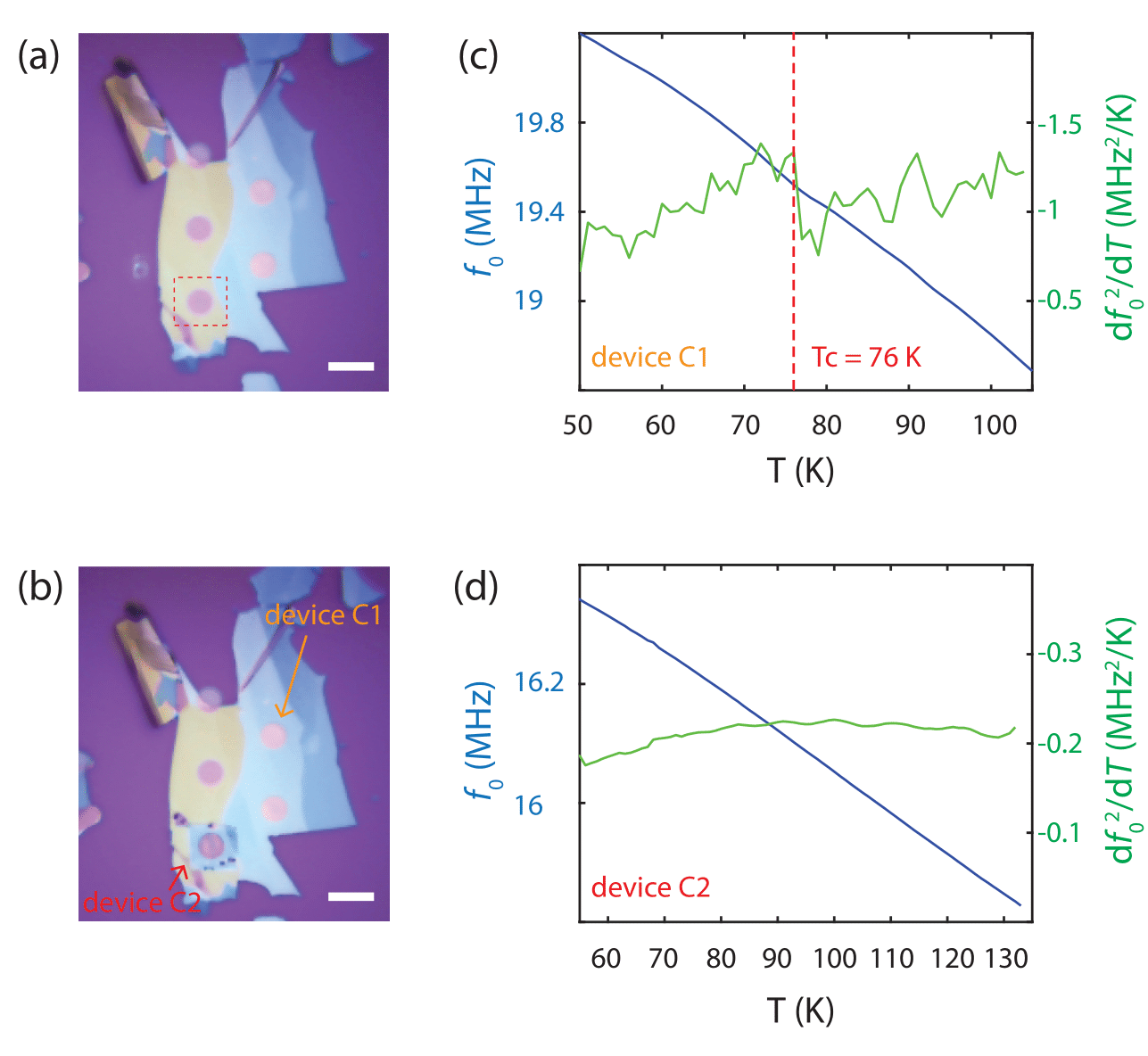}
\caption{Optical image of a 2H-\ch{TaS2} flake on sample C \textbf{(a)} before high intensity laser irradiation and \textbf{(b)} after. The dotted red box in (a) indicates the area before the laser irradiation occurred. The square of discoloration can be seen in (b). Scale bars: 10 \textmu m. \textbf{(c)} Measurement of $f_0$ (blue, left y-axis) and $\frac{d[f^2_0]}{dT}$ (green, right y-axis) of the drum labelled ``device C1''. It displays the expected CDW transition of $T$\textsubscript{CDW} = 76 K. \textbf{(d)} $f_0$ (blue, left y-axis) and $\frac{d[f^2_0]}{dT}$ (green, right y-axis) of the drum labelled ``device C2''. The data shows no signs of a phase transition.}
\label{fig:raman}
\end{figure}

We use the method previously reported by Cartamil et al.\cite{cartamil2015high} who induced oxidation and recrystallization in their 2H-\ch{TaSe2} membrane by shining high intensity laser on it. We use the same protocol in our suspended 2H-\ch{TaS2} flake stamped on substrate C. The optical image of the sample is displayed in Fig. \ref{fig:raman} (a-b). Figure \ref{fig:raman}(a) shows the image of the flake before laser irradiation and (b) after irradiation. The dotted red square is the region where the laser is rastered. As can be seen by the color changes in Fig. \ref{fig:raman}(b), there is an optically observable degradation on the drum labelled ``device C2''. To compare with the pristine state, we performed measurements on the drum labelled ``device C1'' which is of the same flake but has not been irradiated by the laser.

Figure \ref{fig:raman}(c-d) plots $f_0$ (blue, left y-axis) and $\frac{d[f^2_0]}{dT}$ (green, right y-axis) of drums labelled ``device C1'' and ``device C2''. The dotted red line in Fig. \ref{fig:raman}(c) indicates $T$\textsubscript{CDW} = 76 K which is in agreement with literature values of $T$\textsubscript{CDW} in pristine bulk 2H-\ch{TaS2} \cite{sugai1985lattice}. Thus the mechanical response of C1 displays an un-altered CDW transition in 2H-\ch{TaS2}. Figure \ref{fig:raman}(d) shows the same type of data but taken from C2 which has been irradiated with a high intensity laser. There are no clear signs of $T$\textsubscript{CDW}, at least in the temperature ranges considered.

\subsection{Focused ion beam induced disorder on a flake on substrate D}

\begin{figure}[ht]
\includegraphics[width=\columnwidth]{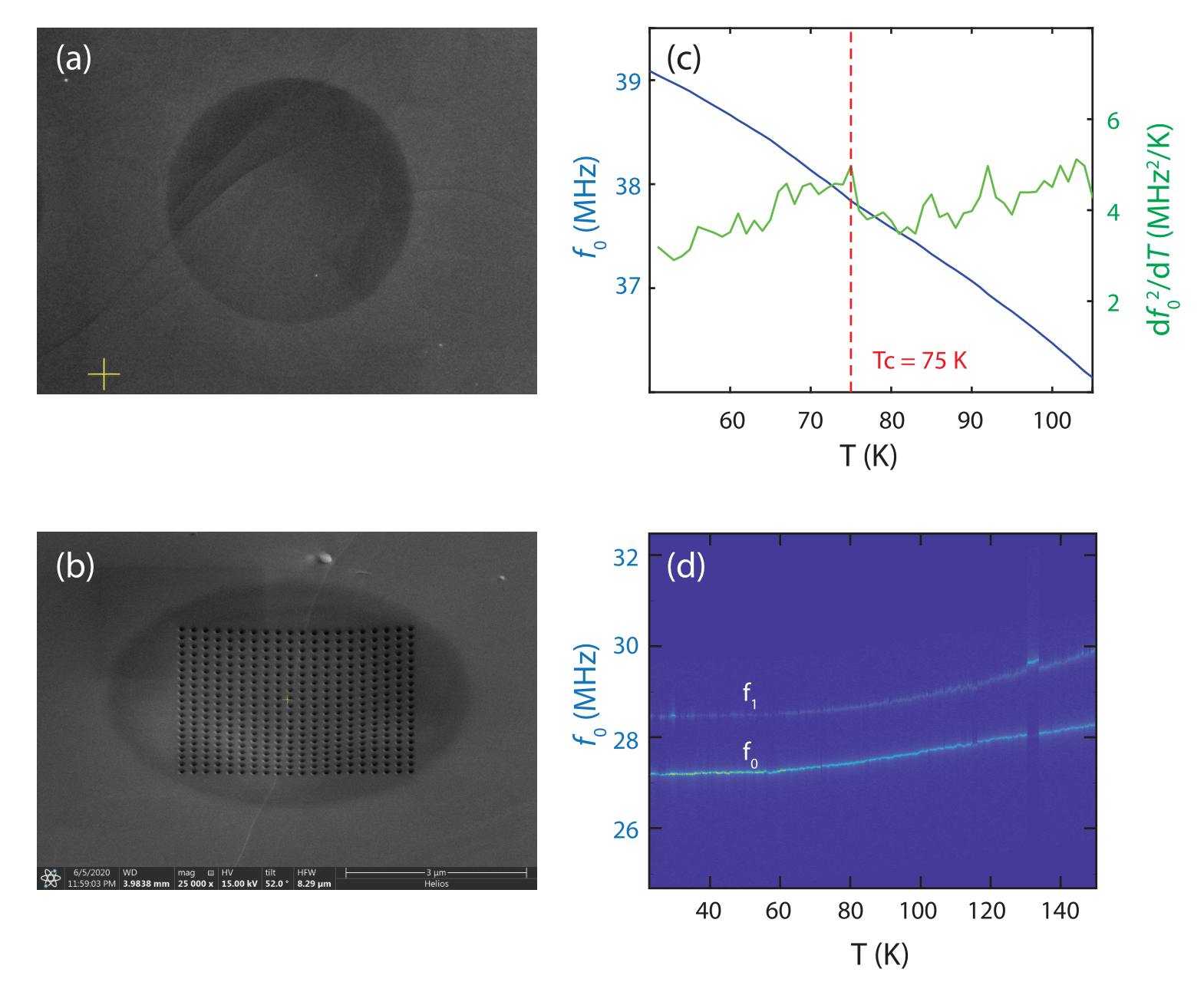}
\caption{\textbf{(a)} Scanning electron microscopy (SEM) image of a pristine 2H-\ch{TaS2} drum (device D1). \textbf{(b)} SEM image taken at an angle of 52$^\circ$, of a drum with 400 pores (device D2) milled by focused ion beam (FIB). \textbf{(c)} Measurement of $f_0$ (blue, left y-axis) and $\frac{d[f^2_0]}{dT}$ (green, right y-axis). Dotted red line indicates the expected $T$\textsubscript{CDW} of 75 K. \textbf{(d)} Measurement of the fundamental frequency, $f_0$, and the first harmonic, $f_1$, represented in a heat map.}
\label{fig:fib}
\end{figure}

In Fig. \ref{fig:fib}(a-b), we show scanning electron microscopy (SEM) images of 2H-\ch{TaS2} flakes stamped on substrate D. Figure \ref{fig:fib}(a) is a SEM image of a pristine flake which has not been irradiated with the Ga+ beam (device D1). Therefore, it can be seen that in the $f_0$ data and $\frac{d[f^2_0]}{dT}$ data of Fig. \ref{fig:fib}(c), there is a CDW transition at 75 K as expected. 

Figure \ref{fig:fib}(b) shows a SEM image of a device (device D2) irradiated with Ga+ ions in a pattern of 20 x 20 array of 50 nm pores that are 200 nm apart. The fundamental mode and the second mode of this FIB irradiated device are plotted in Fig. \ref{fig:fib}(d) in a heat map. As can be seen from the plot, it is striking that the resonance frquencies do not increase with lowering temperature as we have observed in the other devices. Instead, the resonance frequency of this device decreases as the temperature decreases, which signifies that the membrane perhaps has a thermal expansion coefficient with a negative sign possibly arising from the structure \cite{cabras2019micro}. Furthermore, as in the case of the laser irradiated sample in the previous section, there is no observable phase transition in this sample in the range of temperatures we investigated.

\bibliography{enhancedTcdw_supp}
\bibliographystyle{ieeetr}